\documentclass[12pt]{article}
\usepackage[margin=1in]{geometry}
\usepackage{amsmath}
\usepackage{amssymb}
\usepackage{natbib}
\usepackage{amsthm}
\newtheorem{theorem}{Theorem}
\newtheorem{lemma}{Lemma}

\newcommand{\bbeta}{\boldsymbol{\beta}}
\newcommand{\bbetal}{\boldsymbol{\widetilde\beta}^\lambda}
\newcommand{\bbetadl}{\boldsymbol{\widetilde\beta}^{DL}}
\newcommand{\bbetams}{\boldsymbol{\widehat\beta}}
\newcommand{\bSig}{\boldsymbol{\Sigma}}
\newcommand{\bSigHat}{\boldsymbol{\widehat\Sigma}}

\newcommand{\bThetaHat}{\boldsymbol{\widehat\Theta}}
\newcommand{\bIX}{\boldsymbol{X}}
\newcommand{\bIx}{\boldsymbol{x}}
\newcommand{\bx}{\boldsymbol{\widetilde{x}}}
\newcommand{\bb}{\boldsymbol{b}}
\newcommand{\ba}{\boldsymbol{a}}
\newcommand{\Op}[1]{\mathcal{O}_p\left (#1\right )}
\newcommand{\bigO}[1]{\mathcal{O}\left (#1\right )}
\newcommand{\op}[1]{{o}_p\left (#1\right )}
\newcommand{\littleO}[1]{o(#1)}
\newcommand{\norm}[2]{\left \|#1\right \|_{#2}}

\newcommand{\Cone}[2]{\mathcal{C}\left (#1, #2 \right )}
\newcommand{\Slass}{\widehat{S}}
\newcommand{\Isub}{\boldsymbol{\widetilde{I}}}
\newcommand{\bv}{\boldsymbol{v}}
\newcommand{\etavec}{\boldsymbol{\eta}_n}

\DeclareMathOperator*{\argmin}{argmin}

\title{Debiased Lasso After Sample Splitting for Estimation and Inference in High Dimensional Generalized Linear Models}

\author
{Omar Vazquez\thanks{Omar.Vazquez@Pennmedicine.upenn.edu} \\
Department of Biostatistics, Epidemiology and Informatics,\\ University of Pennsylvania, Philadelphia, Pennsylvania, U.S.A.
\and
Bin Nan\thanks{nanb@uci.edu} \\
Department of Statistics, \\University of California, Irvine, California, U.S.A.}

\date{}

\def\spacingset#1{\renewcommand{\baselinestretch}{#1}\small\normalsize}\spacingset{1}

\begin{document}
\maketitle

\begin{abstract}
We consider random sample splitting for estimation and inference in high dimensional generalized linear models, where we first apply the lasso to select a submodel using one subsample and then apply the debiased lasso to fit the selected model using the remaining subsample.
We show that, no matter including a prespecified subset of regression coefficients or not, the debiased lasso estimation of the selected submodel after a single splitting follows a normal distribution asymptotically.  Furthermore, for a set of prespecified regression coefficients, we show that a multiple splitting procedure based on the debiased lasso can address the loss of efficiency associated with sample splitting and produce asymptotically normal estimates under mild conditions.
Our simulation results indicate that using the debiased lasso instead of the standard maximum likelihood estimator in the estimation stage can vastly reduce the bias and variance of the resulting estimates. We illustrate the proposed multiple splitting debiased lasso method with an analysis of the smoking data of the Mid-South Tobacco Case-Control Study.  

\end{abstract}


\spacingset{1.1}

\section{Introduction}
\label{s:intro}

Recent technological advances have allowed biomedical researchers to collect an extraordinary amount of genetic data from patients.
Studies that seek to characterize associations between these genetic factors and medical outcomes face numerous challenges.

Firstly, the standard maximum likelihood estimator (MLE) is not well-defined for classical statistical models such as linear and generalized linear models (GLMs) when the covariates are high dimensional, in particular, $p\gg n$.
Several variable selection and estimation procedures are available for sparse models with high dimensional covariates.
The commonly-used lasso estimator \citep{tibshirani1996regression} introduces a penalty that shrinks many coefficient estimates to exactly zero, thus performing simultaneous variable selection and coefficient estimation.
Similar approaches include the adaptive lasso \citep{zou2006adaptive}, the smoothly clipped absolute deviation (SCAD) estimator \citep{fan2001variable} and elastic net regularization \citep{zou2005regularization}.
Another common approach, which only performs variable selection, is sure independence screening \citep{fan2008sure,fan2010sure}.

Secondly, although penalized regression methods can produce coefficient estimates for GLMs, they typically have substantial biases and do not quantify the uncertainty around these estimates.
Thus they cannot be used to directly construct confidence intervals or to do hypothesis testing in the high dimensional setting.
More recent progress on statistical inference for high dimensional data includes debiased lasso estimators \citep{zhang2014confidence,van2014asymptotically}, MLE-based sample splitting approaches \citep{fei2021estimation,fei2019drawing}, and the decorrelated score test \citep{ning2017general}.
Other work on hypothesis testing without coefficient estimation includes the sample splitting approach for p-values by \citet{meinshausen2009p} and the test developed by \citet{zhu2018linear} for linear models without imposing sparsity assumptions on the regression parameters.

Lastly, we note that some post-model selection inference procedures, such as the simultaneous inference method of \cite{kuchibhotla2020valid} and the sample splitting and bootstrap method of \cite{rinaldo2019bootstrapping}, are applicable to high dimensional covariates, although these in particular only address linear models.
The goal of such methods is distinct from ours.
They are designed to be valid for arbitrary model selection procedures and under potential model misspecification.
This robustness comes at a cost of conservative inference results when a model selection procedure, such as the lasso estimator, can select a superset of the correct model with high probability, which is the setting that we consider below.

Our proposed method is based on sample splitting, a two-stage approach where, first, one part of the data is used to select the covariates to be included in the model, and then the other part is used to estimate the now lower-dimensional model ($p<n$ but $p$ may grow with $n$).
Similar methods have been recently developed for linear models \citep{fei2019drawing,wang2020debiased} and generalized linear models \citep{fei2021estimation} based on the MLE, as well as for Cox proportional hazards models \citep{zhang2022projection} based on the decorrelated score function.
Compared to \cite{fei2021estimation}, we use the lasso instead of sure independence screening for model selection, which allows for more mild theoretical assumptions, as well as the debiased lasso in place of the MLE, which can substantially improve the bias, variance, and confidence interval coverage of the resulting estimators, as shown in our simulations below. 
Note that debiased lasso estimation would be equivalent to using the MLE in sample splitting approaches for linear models.
Our contributions are as follows.

For the model selection component, following \cite{huang2012estimation} we show that, under mild sufficient conditions, the lasso screens out a large portion of the noise variables and selects a model containing the true signal variables with high probability in the case of random design.
Existing work on the selected model size and estimation error of the lasso method for linear regression includes  \citet{bickel}, \citet{zhang2008sparsity}, \citet{meinshausen2009lasso}, and \citet{zhao2006model}, under differing assumptions on the covariates.
Similar results were shown for generalized linear models by \citet{van2008high} and \cite{huang2012estimation}, only the latter of which addressed model selection consistency and the number of selected noise variables. 
Their results imply that the lasso selected model size is of the same order as the number of nonzero regression coefficients, with high probability. 
This is essential because if the selected model size is too large then its design matrix may not have full rank, even asymptotically, and thus the regression coefficients will still not be estimable via MLE or even the refined debiased lasso of \citet{xia2021debiased}.

For the lower-dimensional model estimation stage, while a naive approach would be to use the standard maximum likelihood estimator with the selected covariates, which is commonly used in sample splitting literature, we instead apply the recently developed refined debiased lasso approach of \citet{xia2021debiased}.
This method is better suited for models that contain a substantial number of noise variables, as is typically expected after variable selection.
This is a more desirable approach because the conditions for the GLM lasso to not select any noise variables, for example, are known to be quite stringent \citep{huang2012estimation}.
We illustrate the potentially large difference in performance through simulations, where the MLE exhibits a strong bias that inflates the size of estimated coefficients, and this bias increases with the true signal strength, in contrast to the approximately unbiased estimates from the debiased lasso.
Such tendencies are discussed further by \citet{xia2021debiased} in the lower-dimensional setting.

For a set of prespecified coefficients, since their estimators based on a single sample split suffer from a loss of efficiency due to only using part of the data for estimation, we further investigate the idea of averaging estimates across multiple sample splits.
\cite{fei2019drawing} proposed a multiple splitting procedure for linear models, where they required model selection consistency for their asymptotic theory.
\cite{wang2020debiased} discussed multiple splitting for a single treatment variable in linear models, under much milder assumptions.
Recently, \cite{fei2021estimation} proposed a multiple splitting procedure for generalized linear models based on the MLE under a partial orthogonality condition, which requires that the signal variables be independent of the noise variables.
For our proposed multiple splitting procedure, we apply the debiased lasso estimator instead of the MLE in the estimation stage, and show through simulations that our procedure results in approximately unbiased estimates and substantially reduced variability compared to the MLE-based multiple splitting that is often biased. For the theoretical development, we adapt the mild conditions of \cite{wang2020debiased} to GLMs and show the asymptotic normality of our proposed approach.
As evidenced by simulations, our multiple splitting estimator can produce confidence intervals with the nominal coverage.

\section{Methods}
\label{s:method}

We first provide brief overviews of lasso and debiased lasso methods, then introduce our proposed sample splitting debiased lasso methods for GLMs. 

\subsection{Notation}
\label{s:notation}

For a positive integer $p$, we denote the size of any index set $S\subseteq\{1,\ldots,p\}$ by $|S|$.
For a $p$-dimensional vector $\boldsymbol{b}$ and a $p\times p$ dimensional matrix $\boldsymbol{B}$, let $\boldsymbol{b}_S$ denote the $|S|$-dimensional subvector with entries indexed by $S$, and similarly $\boldsymbol{B}_S$ denote the $|S|\times |S|$ submatrix with rows and columns indexed by $S$.
The $\ell_q$ norm is denoted as $\norm{\boldsymbol{b}}{q}$ for $q\geq 1$.
For positive sequences $a_n$ and $b_n$, we write $a_n=\bigO{b_n}$ if there exists a constant $c>0$ and $N>0$ such that $a_n/b_n<c$ for all $n>N$, we write $a_n=\littleO{b_n}$ if $a_n/b_n\rightarrow 0$ as $n\rightarrow\infty$, and we write $a_n\sim b_n$ if $a_n=\bigO{b_n}$ and $b_n=\bigO{a_n}$.

Let $(y_i, \bx_i)$, $i=1,\ldots,n$, be independent and identically distributed copies of $(y, \bx)$, where $y$ is a scalar-valued response variable and $\bx$ is a $p$-dimensional vector of covariates.
We consider the high dimensional setting where $p$ can be much larger than $n$.
Let $\bIx_i = (1,\bx_i^T)^T$ and denote the $n\times (p+1)$ design matrix by $\bIX$, with $i$th row $\bIx_i^T$ and $j$th column $\bIX_j$. 
We assume without loss of generality that $\bx$ has mean zero.
For any function $f(y,\bIx)$, we define $P_nf = n^{-1}\sum_{i=1}^n f(y_i, \bIx_i)$.

We consider generalized linear models \citep{mccullagh2019generalized} with canonical link function and known dispersion parameter.
Denote
\[
    \rho_{\bbeta}(y, \bIx) 
    = \rho(y, \bIx^T\bbeta)
    = A(\bIx^T\bbeta) - y\bIx^T\bbeta
\]
for a known twice-differentiable function $A(\cdot)$, and its first and second derivatives with respect to $\bbeta$ as $\dot{\rho}_{\bbeta}$ and $\ddot{\rho}_{\bbeta}$, respectively.
The negative log-likelihood for $\bbeta$ is then $P_n\rho_{\bbeta} = n^{-1}\sum_{i=1}^n \rho_{\bbeta}(y_i, \bIx_i)$ with score function $P_n\dot{\rho}_{\bbeta}=n^{-1}\sum_{i=1}^n  \{A'(\bIx_i^T\bbeta) - y_i \}\bIx_i $ and Hessian matrix $P_n\ddot{\rho}_{\bbeta}(y_i, \bIx_i) = n^{-1}\sum_{i=1}^n A''(\bIx_i^T\bbeta)\bIx_i\bIx_i^T$. 
Denote $\bSig_{\bbeta} = E\{P_n\ddot{\rho}_{\bbeta}(y_i, \bIx_i)\}$.
The $(p+1)$-dimensional unknown true coefficient vector $\bbeta^0$ is assumed to be sparse, and we denote the index set of signal variables by $S_0 = \{j:\bbeta_j^0\neq 0\}$, with $s_0=|S_0|$.
The quantities $p$, $\bbeta^0$, $s_0$, and $S_0$ are allowed to change with $n$.
For practical applications, one would generally consider $s_0$ and $\bbeta^0_{S_0}$ to be fixed so that the regression coefficients have the usual interpretation in terms of the conditional mean $E[y|\bIx]$ as $p$ grows with $n$. Letting $s_0$ grow with $n$ does not maintain the log odds ratio interpretation of $\bbeta^0$, for example, in a logistic regression model.  

\subsection{The lasso estimator for generalized linear models}
\label{s:lasso}

The lasso estimator $\bbetal$ for the GLM parameters $\bbeta^0$ is, given a tuning parameter $\lambda>0$, a minimizer of the penalized negative log likelihood 
\[
  \bbetal = \argmin_{\bbeta}  P_n\rho_{\bbeta} + \lambda\norm{\bbeta}{1}.
\]
In general there is no closed form solution for $\bbetal$, but the objective function is convex and can be optimized efficiently with widely available software \citep{friedman2010regularization}.
The penalization shrinks the estimated coefficients towards zero, with many of their values set exactly to zero, resulting in a sparse estimator $\bbetal$.
Model selection may then be performed by keeping only the covariates with nonzero estimated coefficients.
In practice, a small subset with a finite number of coefficients may be left unpenalized so that they are never excluded from the selected model.
In our simulations and data analysis, we do not penalize the intercept.

\subsection{The debiased lasso estimator}
\label{s:debiased}

\cite{zhang2014confidence} and \citet{van2014asymptotically} proposed debiased, also called desparsified, lasso estimators for linear models and GLMs, respectively.
This procedure may be used to obtain coefficient estimates and confidence intervals across the entire parameter vector $\bbeta^0$ in high dimensional settings.
It does not rely on sample splitting or variable selection.
Instead, using the lasso estimator $\bbetal$ as an initial value, it performs a single Newton iteration for minimizing the negative log likelihood $P_n\rho_{\bbeta}$, using an approximate inverse for the Hessian matrix $P_n\ddot{\rho}_{\bbetal}$ such as a nodewise lasso estimator.
The resulting desparsified lasso estimator is no longer sparse and is, under further conditions, asymptotically unbiased with a normal distribution.

A key assumption for accurately estimating  the inverse of the Hessian matrix in high dimensional settings, however, is the sparsity assumption on $\bSig_{\bbeta^0}^{-1} = [E\{A''(\bIx^T\bbeta^0)\bIx\bIx^T\}]^{-1}$.
For generalized linear models, each entry of this matrix generally depends on all the signal variables due to the non-identity link function, so the sparsity assumption is difficult to interpret and often may not hold in practice.

Such issues were discussed further by \citet{xia2021debiased}, who proposed a refined debiased lasso method for lower dimensional GLMs with a diverging number of covariates.
In that setting and under some standard conditions, $\bSigHat_{\bbetal} = P_n\ddot{\rho}_{\bbetal}$ is directly invertible with probability going to one, and the resulting estimator 
\begin{equation}\label{dlasso}
    \bbetadl = \bbetal - \bThetaHat_{\bbetal} P_n \dot{\rho}_{\bbetal}
\end{equation}
with estimated covariance matrix $\bThetaHat_{\bbetal}/n = \bSigHat_{\bbetal}^{-1}/n$
was shown to be asymptotically normal.
Additionally, it outperforms both the original desparsified lasso estimator and the MLE in several simulation settings of lower-dimensional GLMs with sparse regression coefficients.
It can also be computed using standard GLM software packages for the lasso and MLE. 

\subsection{The debiased lasso after single sample splitting}
\label{s:sample-splitting}

We propose a debiased lasso after sample splitting procedure that uses the lasso for model selection and the refined debiased lasso for estimation and inference.
For a fixed splitting proportion $q\in (0,1)$, we randomly split the sample $(y_i,\bIx_i)_{i=1}^n$ into two subsets: $\mathcal{D}_1$ containing $n_1= qn$ individuals to be used for variable selection, and $\mathcal{D}_2$ containing $n_2 = n-n_1$ individuals to be used for estimation.
For a pre-specified fixed set of coefficient indices $S$ that is allowed to be an empty set, do the following:

\begin{enumerate}
    \item \label{ss-selection} using the subsample $\mathcal{D}_1$, fit a GLM lasso estimator $\bbetal$ to obtain the selected model $\Slass = S\cup \{j:\bbetal_j\neq 0\}$;
    
    \item\label{ss-estimation}
    using the subsample $\mathcal{D}_2$, compute the debiased lasso estimator $\bbetadl_{\Slass}$ for $\bbeta^0_{\Slass}$ based on a GLM with covariates $\Slass$, and use the estimated covariance matrix of $\bbetadl_{\Slass}$, $\bThetaHat_{\bbetal,\Slass}/n_2$, to construct $(1-\alpha)$\% confidence intervals for any contrast $\ba_{\Slass}^T\bbeta^0_{\Slass}$ based on a normal distribution.
\end{enumerate}

Because the two subsamples are independent, we can make asymptotically valid statistical inference on any covariate in $\Slass$ as long as this set contains the true signal variables when $n$ is large enough.
In practice, covariates with small coefficients may not be selected in the first stage, so making conclusions about the statistical significance of omitted variables from step \ref{ss-selection} is not advised.
The large bias that can potentially occur when $S$ is a null set is illustrated through simulations in Section \ref{s:simulations}.
Furthermore, although the size of the fitted model $\Slass$ must be low-dimensional, this does not prevent one from obtaining estimates for each individual coefficient.
For example, step \ref{ss-estimation} may be repeated for each individual coefficient, i.e. for $S=\{1\}, \{2\},\ldots, \{p\}$, along with an appropriate correction for multiple testing when computing p-values and confidence intervals.

\subsection{The debiased lasso after multiple sample splitting}
\label{s:sample-splitting-m}

As previously noted, the single sample splitting estimator suffers from a loss of efficiency due to only using a fraction of the total sample to estimate $\bbeta^0$.
The following multiple splitting procedure addresses this issue. For a pre-specified fixed set of coefficient indices $S$,
we now generate $B$ random splits of the sample $(y_i,\bIx_i)_{i=1}^n$, denoted $(\mathcal{D}_{1,b}, \mathcal{D}_{2,b})$ for $b=1,\ldots,B$, and do the following:

\begin{enumerate}
    \item for $b=1,\ldots,B$,
    \begin{enumerate}
        \item using the subsample $\mathcal{D}_{1,b}$, fit a GLM lasso estimator $\bbetal_b$ to obtain the selected model $\Slass_b = S\cup \{j:\bbetal_{b,j}\neq 0\}$;
        
        \item\label{ms-estimation}  using the subsample $\mathcal{D}_{2,b}$, compute the debiased lasso estimator $\bbetadl_{S,b}$ for $\bbeta^0_{S}$ based on a GLM with covariates $\Slass_b$;
    \end{enumerate}
    
    \item finally the multiple splitting estimators are $\bbetams_S = B^{-1}\sum_{b=1}^B \bbetadl_{S,b}$.
\end{enumerate}

Note that all target coefficients to estimate must be pre-specified, in contrast to the single split procedure.
For exploratory analysis, where step \ref{ms-estimation} may be repeated for each individual coefficient, the multiple splitting estimator typically has a substantially more strict significance threshold compared to single split inference on $\Slass$ due to correcting for $p$, $p>|\Slass|$, multiple comparisons.

We apply the same variance estimator as \cite{fei2021estimation}, based on the nonparametric delta method with a bias correction for controlling the Monte Carlo error \citep{efron2014estimation,wager2014confidence}.
Letting $\bv_b$ denote the $n$-vector of sampling indicators such that $\bv_{b,i}\in \{0,1\}$ is equal to one if $i\in \mathcal{D}_{2,b}$, and is zero otherwise, 
\[
  \widehat{Var}\left (\bbetams_j \right ) 
  = \widehat{V}_j - \frac{n(n_2)}{B^2(n-n_2)}\sum_{b=1}^B \left (\bbetadl_{j,b} - \bbetams_j\right )^2,
\]
\[
  \widehat{V}_j = \frac{n(n-1)}{(n-n_2)^2}\sum_{i=1}^n \left \{\frac{1}{B}\sum_{b=1}^B \left (\bv_{b,i} - \bar{\bv}_i\right )(\bbetadl_{j,b} - \bbetams_j) \right \}^2,
\]
where $\bar{\bv}_i = B^{-1}\sum_{b=1}^B \bv_{b,i}$, and $(1-\alpha)$\% confidence intervals may be constructed based on a normal approximation.
Similarly, the estimated covariance matrix is
\[
  \widehat{Var}\left (\bbetams_S \right ) 
  = \frac{n(n-1)}{(n-n_2)^2}\sum_{i=1}^n \boldsymbol{\widehat{C}}_i\boldsymbol{\widehat{C}}_i^T
  -  \frac{n(n_2)}{B^2(n-n_2)}\sum_{b=1}^B \boldsymbol{\widehat{D}}_b\boldsymbol{\widehat{D}}_b^T
\]
\[
  \boldsymbol{\widehat{C}}_i = \frac{1}{B}\sum_{b=1}^B \left (\bv_{b,i} - \bar{\bv}_i\right )\left (\bbetadl_{S,b} - \bbetams_S\right ), \
  \boldsymbol{\widehat{D}}_b = \bbetadl_{S,b} - \bbetams_S.
\]

Both single and multiple sample splitting procedures require that the matrices $\bSigHat_{\bbetal,\Slass}$ computed from each estimation split are invertible.
One way to encourage this is to set a hard limit for the selected model size, which would set a lower bound for the parameter $\lambda$ in the lasso.
In our simulations and real data analysis we do not restrict $\lambda$, which is chosen by cross validation, and instead allow the glm function in R to follow its default behavior of automatically dropping covariates from the model when the design matrix is ill-conditioned.

As the number of random splits $B$ grows we typically have diminishing gains in efficiency relative to the increased computational cost.
Although our theoretical results consider the maximum possible value of $B={n\choose n_2}$, our simulation results indicate that $B=1,000$ is sufficient for many practical settings with up to several hundreds of individuals in the sample.

\section{Theoretical Results}
\label{s:theory}

For the single split estimator, we make the following assumptions:

\begin{enumerate}
    \item\label{cond-subg} The covariates $\bIx_i$ are sub-Gaussian random vectors and $\norm{\bIX}{\infty}\leq K$ almost surely for some constant $K>0$.
    
    \item\label{cond-eig} $\bSig_{\bbeta^0}$ and $E[\bIX^T\bIX/n]$ are positive definite with eigenvalues bounded from above and away from zero.
    
    \item\label{cond-lip} The derivatives $\dot{\rho}(y,a) = \partial\rho(y,a)/\partial a$ and $\ddot{\rho}(y,a) = \partial^2\rho(y,a)/\partial a^2$ exist for all $(y, a)$ and there exists a $\delta>0$ and constant $c_{Lip}>0$ such that $\ddot{\rho}$ is locally Lipschitz
    \[
      \max_{a_0\in \{\bIx_i^T\bbeta^0\}} 
      \sup_{\max\left (|a-a_0|, |\widehat{a} - a_0|\right )\leq\delta} 
      \sup_{y\in\mathcal{Y}} 
      \frac{|\ddot{\rho}(y,a) - \ddot{\rho}(y, \widehat{a})|}{|a-\widehat{a}|}\leq c_{Lip}.
    \]
    Also, there exist constants $K_1,K_2>0$ such that the derivatives are bounded:
    \[
      \max_{a_0\in \{\bIx_i^T\bbeta^0\}} 
      \sup_{y\in\mathcal{Y}}
      |\dot{\rho}(y,a_0)|\leq K_1,
    \]
    \[
      \max_{a_0\in \{\bIx_i^T\bbeta^0\}} 
      \sup_{y\in\mathcal{Y}} 
      |\ddot{\rho}(y,a_0)|\leq K_2.
    \]
    
    \item\label{cond-linpred} $\norm{\bIX\bbeta^0}{\infty}$ is bounded above almost surely.
    
    \item\label{cond-betamin} The sparsity $s_0$ satisfies $s_0\log(p)/\sqrt{n} = \littleO{1}$ and the regression parameters are large enough that $\min_{j\in S_0}|\bbeta_j^0| \geq \bigO{s_0\sqrt{\log(p)/n}}$ as $n\rightarrow\infty$. 
\end{enumerate}

Assumptions \ref{cond-subg}, \ref{cond-eig}, and \ref{cond-linpred} are typical in high dimensional literature.
Bounded covariates are common in practice, including dummy variables for categorical covariates, minor allele counts, and physical measurements.
Assumption \ref{cond-lip} is required in order to apply the results for the refined debiased lasso estimator of \cite{xia2021debiased}.
All but the boundedness condition on $\dot{\rho}$ are satisfied when the function $A$ is twice continuously differentiable, as is the case with commonly-used GLMs, due to assumption \ref{cond-linpred}.
The derivative $\dot{\rho}$ is also bounded if the response variable has bounded support, such as in logistic regression, and this condition can be relaxed to include other common GLMs by instead assuming sub-exponential tails, as in Lemma 3.4 of \cite{ning2017general}. 
Assumption \ref{cond-betamin} guarantees that, with high probability, model selection based on the lasso does not miss any signal variables, so that the selected model is not misspecified.
This is a standard condition for sample splitting procedures that is not required by the desparsified lasso \citep{van2014asymptotically}.

For multiple splitting, we make the following additional assumptions for the set of target covariates $S$, where $\ba$ is a $(p+1)$-vector such that $\norm{\ba_S}{2}=1$ and $\norm{\ba_{S^c}}{2}=0$.
Let $\mathcal{Z} = (y_i,\bIx_i)_{i=1}^n$ denote the entire data set.

\begin{enumerate}
    \setcounter{enumi}{5}
    \item\label{cond-ms1} For independent sampling indicator vectors $\bv$ and $\boldsymbol{\widetilde{v}}$ with corresponding fitted models $\Slass$ and $\widetilde{S}$, define
    \[h_{i,n} 
    = \left [E\left (\bv_i\ba_{\Slass}^T\bSig_{\bbeta^0, \Slass}^{-1}\Isub_{\Slass}\Big |\mathcal{Z} \right )
    - E\left (\bv_i\ba_{\widetilde{S}}^T\bSig_{\bbeta^0, \widetilde{S}}^{-1}\Isub_{\widetilde{S}}\Big |\mathcal{Z} \right )\right ]\dot{\rho}_{\bbeta^0}(y_i,\bIx_i) \]
    where $\Isub_S$ is the $|S|\times (p+1)$ matrix such that $\Isub_S\bb = \bb_S$ for any $(p+1)$-vector $\bb$, and the expectations are with respect to the sampling weights, conditional on the data.
    We assume that $\sum_{i=1}^n h_{i,n}/\sqrt{n} =\littleO{1}$ and that $P_n\ddot{\rho}_{\bbetal}$ is invertible when computed from any estimation subsample.
    \item\label{cond-ms2} There exists a random $(p+1)$-vector $\etavec$ independent of the data such that $\norm{\etavec}{\infty}$ is bounded and 
    \[\norm{E\left (\ba_{\Slass}^T\bSig_{\bbeta^0, \Slass}^{-1}\Isub_{\Slass}\Big |\mathcal{Z} \right ) - \etavec^T}{1} = \op{1/\sqrt{\log(p)}}. \]
\end{enumerate}

Similar conditions were used and discussed further by \cite{wang2020debiased} in the context of linear models.
Since, for low-dimensional GLMs, the refined debiased lasso estimator is asymptotically equivalent to the MLE $\widetilde{\bbeta}^{ML}$ in the sense that $\sqrt{n}(\bbetadl_j - \widetilde{\bbeta}_j^{ML})=\op{1}$ for all $j$, our theoretical arguments also apply to multiple splitting with the MLE, using the lasso for model selection.
\cite{fei2021estimation} used a stronger partial orthogonality condition, where the signal variables are independent of the noise variables, to derive the asymptotic normality of their MLE-based multiple splitting estimator.
We further discuss the motivation behind Assumptions \ref{cond-ms1} and \ref{cond-ms2} and their relationship to the assumptions used by other work on sample splitting procedures in \ref{appendix:disc}, but provide a brief overview here.

Under Assumptions \ref{cond-subg}-\ref{cond-betamin} and the sparsity rates required in Theorem \ref{thm-ms} below, it can be shown that
\[
  \sqrt{n}\ba_S^T(\bbetams_S - \bbeta_S^0)
  = \frac{(1-q)^{-1}}{\sqrt{n}}\sum_{i= 1}^n E\left [\ba_{\Slass}^T\bSig_{\bbeta^0, \Slass}^{-1}\Isub_{\Slass}   \bv_{i} | \mathcal{Z}\right ]\dot{\rho}_{\bbeta^0}(y_i, \bIx_i) + \op{1}.
\]
Therefore, in order to prove asymptotic normality we need to control the randomness of the vector $\ba_{\Slass}^T\bSig_{\bbeta^0, \Slass}^{-1}\Isub_{\Slass}\bv_{i}$ averaged across all sample splits that exclude the $i$th sample from model selection (i.e. $\bv_i=1$).
The ideal case, when this average is deterministic, can be achieved under model selection consistency, so that $\Slass$ is fixed at $S\cup S_0$ for each sample split, or if the set of covariates that are always included in the fitted model are independent of all remaining covariates, essentially giving $\bSig_{\bbeta^0}^{-1}$ a block diagonal structure.
Either of these two conditions imply that our less restrictive assumptions hold, where we only require that the effect of always excluding a single sample from model selection is asymptotically negligible (Assumption \ref{cond-ms1}) and that the average converges in probability to a bounded random vector with moderate rate (Assumption \ref{cond-ms2}).

The variable screening properties and asymptotic normality of our estimators are presented in the following three theorems, with their proofs given in \ref{appendix:thm1} through \ref{appendix:thm3}, respectively.
Note that any variable selection procedure with the same screening properties as the lasso may be used in our sample splitting procedures.

\begin{theorem}\label{thm-lasso}
    (Variable screening properties for the GLM lasso.) 
    For a choice of tuning parameter $\lambda\sim \sqrt{\log(p)/n}$ and under assumptions \ref{cond-subg}-\ref{cond-betamin}, the lasso estimator $\bbetal$ and its selected model $\Slass = \{j:\widetilde{\beta}_j^\lambda\neq 0\}$ satisfy
    \[
       P\left (\Slass\supseteq S_0, |\Slass|\leq ks_0 \right )\geq 1 - c_1/p^{c_2} 
    \]
    for some positive constants $k, c_1, c_2$.
\end{theorem}

\begin{theorem}\label{thm-ss}
    (Asymptotic normality for single sample splitting.) For the single split debiased lasso estimator $\bbetadl_{\Slass}$ with both lasso tuning parameters of the order $\sqrt{\log(p)/n}$,  $s_0\log(s_0)\sqrt{s_0/n} = \littleO{1}$, $|S|=\bigO{s_0}$, and under assumptions \ref{cond-subg}-\ref{cond-betamin},
    $\bSigHat_{\bbetal, \Slass}$ is invertible with probability going to one, and for any $(p+1)$-vector $\ba$ such that $\norm{\ba}{2} = \norm{\ba_{\Slass}}{2}=1$, we have
    \[
      \frac{\sqrt{n_2}\ba_{\Slass}^T(\bbetadl_{\Slass} - \bbeta^0_{\Slass})}{(\ba_{\Slass}^T\bThetaHat_{\bbetal,\Slass}\ba_{\Slass})^{1/2}}\rightarrow N(0,1)
    \]
   in distribution as $n\rightarrow\infty$.
\end{theorem}

\begin{theorem}\label{thm-ms}
    (Asymptotic normality for multiple sample splitting.) For the multiple splitting debiased lasso estimator $\bbetams_S$ with all lasso tuning parameters of the order $\sqrt{\log(p)/n}$, $s_0\log(s_0)\sqrt{s_0/n} = \littleO{1}$, $|S|=\bigO{s_0}$, and under assumptions \ref{cond-subg}-\ref{cond-ms2},
    for any $(p+1)$-vector $\ba$ such that $\norm{\ba}{2} = \norm{\ba_{S}}{2}=1$, we have
    \[
      \frac{\sqrt{n}\ba_S^T(\bbetams_S - \bbeta^0_S)}{(\etavec^T\bSig_{\bbeta^0}\etavec)^{1/2}}\rightarrow N(0,1)
    \]
    in distribution as $n\rightarrow\infty$.
\end{theorem}

\section{Simulations}
\label{s:simulations}

We apply the proposed estimating methods with sample splitting in several simulations settings with high dimensional covariates in order to assess the potential benefits of using the lasso for model selection and the debiased lasso in place of the MLE for making statistical inference, and of averaging estimates across multiple sample splits (MS) as opposed to using a single split (SS) estimator for a fixed set of coefficients.
Simulation results are all from logistic regression models, where the benefits of using the debiased lasso are particularly apparent.
This may be due to the numerical instability of the Hessian matrix of the negative log-likelihood $P_n\ddot{\rho}_{\bbeta}(y_i, \bIx_i) = n^{-1}\sum_{i=1}^n \hat{p}(\bIx_i^T\bbeta)\left [1-\hat{p}(\bIx_i^T\bbeta)\right ]\bIx_i\bIx_i^T$, where $\hat{p}(\cdot)$ is the cdf of the standard logistic distribution.
Note $\hat{p}(\bIx_i^T\bbeta)$ will be close to zero or one for large coefficient values, which can result in a near-singular Hessian matrix.
Therefore the debiased lasso approach of performing a single-step maximization of the log-likelihood starting from an initial lasso estimator that is biased towards zero can help alleviate this issue.  

In all simulations, $n=500$ samples are generated from a logistic regression model with $p=700$ covariates.
The covariates are generated from a $N(\boldsymbol{0}, \boldsymbol{\Sigma})$ distribution before being truncated at $\pm 3$. 
The covariance matrix $\boldsymbol\Sigma$ has ones on the diagonal and either an AR(1) correlation structure $\boldsymbol\Sigma_{jk} = 0.5^{|j-k|}$ or a compound symmetry structure $\boldsymbol\Sigma_{jk} = 0.5$ $(j\neq k)$.
Further simulations that demonstrate the robustness of our procedures under higher covariate correlations are presented in \ref{appendix:sims}.
There are $s_0=6$ signal covariates, and their index set was randomly chosen.
The corresponding coefficient values are $\bbeta^0_{S_0} = (-1.5, -1, -0.5, 0.5, 1, 1.5)^T$.

We consider all signal covariates together with two randomly chosen noise covariates. For each coefficient $\beta_j$,  we assess the single and multiple splitting debiased lasso estimators fit on $\Slass\cup\{j\}$, as well as the corresponding MLE-based estimators.
The splitting proportion is $q=0.5$.
We also provide results from the oracle MLE, fit on $S_0\cup \{j\}$, for reference, using either the entire sample or half of the sample.
Model selection is performed using lasso estimators with the values for $\lambda$ chosen by 10-fold cross validation.
The multiple splitting estimators use $B=1,000$ splits.
We use the R package glmnet for lasso estimation and the built-in glm function for MLE and debiased lasso estimation, using the control and start arguments to specify a single iteration starting at an initial lasso estimate for the latter.

The simulation results for the AR(1) correlation structure are summarized in Table \ref{table-ar1-5}.
The single split MLE exhibits a bias that greatly inflates the size of the estimated coefficients, and the magnitude of this bias increases with the signal strength.
In contrast, the single split debiased lasso estimator is approximately unbiased.
Averaging across multiple splits does not improve the bias of the MLE, which is also apparent to a lesser extent in the logistic regression simulations of \cite{fei2021estimation} summarized in their Table 3.
The MLE has substantially higher variability than the debiased lasso, even for the multiple splitting estimators.
The 95\% confidence interval coverage for noise variables is roughly the same across all considered methods.
For signal variables, however, the MLE has poor coverage that actually worsens after multiple splitting.
This issue appears to be particularly severe in logistic regression models and is mild in some other GLMs such as Poisson regression.
In contrast, the debiased lasso after multiple splitting performs well in achieving the nominal 95\% coverage for all considered coefficients.
All single split standard errors tend to be underestimated for signal variables, leading to slight undercoverage, while the multiple splitting standard errors are approximately unbiased.
Multiple splitting drastically lowers the variability of estimates from either the MLE or debiased lasso compared to a single split.
This produces a dramatic improvement in rejection rate for small coefficients.
Note that the rejection rates for the MLE estimators are inflated due to their bias, which partially offsets the wider confidence interval length.
In summary, for each pre-specified coefficient, the multiple splitting debiased lasso estimator provides the best performance in terms of bias, variability, and confidence interval coverage in this simulation setting, where the correlation between covariates decays rather quickly.

\begin{table}
\caption{Logistic regression simulation results for $n=500$, $p=700$, $s_0=6$, and AR(1) correlation structure with parameter 0.5. Selection results refer to the lasso in the single split estimator, where the average selected model size was 41. The fitted model for estimating each $\beta_j$ was $\Slass\cup\{j\}$. Nominal confidence interval coverage probabilities are 0.95.} \label{table-ar1-5}
{\begin{tabular*}{\textwidth}{l l r r r r r r r r}
\hline  & Estimator & $\beta_{489}$ & $\beta_{130}$ & $\beta_{680}$ & $\beta_{488}$ & $\beta_{476}$ & $\beta_{190}$ & $\beta_{510}$ & $\beta_{336}$\\ \hline
$\bbeta^0_j$ &  & -1.50 & -1.00 & -0.50 & 0.00 & 0.00 & 0.50 & 1.00 & 1.50\\
 Selection Rate &  & 1.00 & 1.00 & 0.64 & 0.13 & 0.07 & 0.69 & 1.00 & 1.00\\ \hline
 Bias & Debiased SS & 0.01 & 0.06 & 0.00 & -0.04 & -0.01 & -0.03 & 0.00 & 0.00\\
  & MLE SS & -0.52 & -0.30 & -0.24 & -0.06 & -0.01 & 0.18 & 0.37 & 0.54\\
  & Debiased MS & 0.01 & 0.04 & 0.02 & -0.02 & 0.00 & -0.02 & -0.01 & 0.00\\
  & MLE MS & -0.59 & -0.37 & -0.21 & -0.02 & -0.01 & 0.21 & 0.40 & 0.60\\
  & Oracle $(n_2)$ & -0.09 & -0.05 & 0.00 & -0.01 & 0.00 & 0.04 & 0.07 & 0.09\\
  & Oracle $(n)$ & -0.05 & -0.01 & -0.01 & -0.02 & 0.00 & 0.01 & 0.04 & 0.05\\ \hline
 Coverage & Debiased SS & 0.84 & 0.89 & 0.92 & 0.94 & 0.95 & 0.88 & 0.87 & 0.89\\
  & MLE SS & 0.73 & 0.86 & 0.84 & 0.90 & 0.92 & 0.86 & 0.77 & 0.69\\
  & Debiased MS & 0.94 & 0.94 & 0.93 & 0.95 & 0.97 & 0.96 & 0.96 & 0.95\\
  & MLE MS & 0.39 & 0.64 & 0.81 & 0.94 & 0.94 & 0.79 & 0.58 & 0.40\\
  & Oracle $(n_2)$ & 0.95 & 0.96 & 0.94 & 0.95 & 0.96 & 0.94 & 0.96 & 0.97\\
  & Oracle $(n)$ & 0.96 & 0.95 & 0.94 & 0.92 & 0.94 & 0.95 & 0.96 & 0.96\\ \hline
 Rejection Rate & Debiased SS & 1.00 & 0.99 & 0.74 & 0.06 & 0.05 & 0.62 & 1.00 & 1.00\\
  ($H_0: \beta_j=0$) & MLE SS & 1.00 & 0.99 & 0.77 & 0.10 & 0.08 & 0.71 & 1.00 & 1.00\\
  & Debiased MS & 1.00 & 1.00 & 0.92 & 0.05 & 0.03 & 0.94 & 1.00 & 1.00\\
  & MLE MS & 1.00 & 1.00 & 0.94 & 0.06 & 0.06 & 0.95 & 1.00 & 1.00\\
  & Oracle $(n_2)$ & 1.00 & 1.00 & 0.74 & 0.05 & 0.04 & 0.80 & 1.00 & 1.00\\
  & Oracle $(n)$ & 1.00 & 1.00 & 0.97 & 0.09 & 0.06 & 0.98 & 1.00 & 1.00\\ \hline
 Standard Error & Debiased SS & 0.22 & 0.20 & 0.19 & 0.21 & 0.18 & 0.19 & 0.20 & 0.22\\
  & MLE SS & 0.37 & 0.30 & 0.28 & 0.29 & 0.25 & 0.27 & 0.31 & 0.37\\
  & Debiased MS & 0.18 & 0.15 & 0.14 & 0.14 & 0.12 & 0.14 & 0.16 & 0.18\\
  & MLE MS & 0.27 & 0.22 & 0.20 & 0.22 & 0.19 & 0.20 & 0.22 & 0.27\\
  & Oracle $(n_2)$ & 0.26 & 0.22 & 0.20 & 0.22 & 0.19 & 0.20 & 0.22 & 0.26\\
  & Oracle $(n)$ & 0.18 & 0.15 & 0.14 & 0.15 & 0.13 & 0.14 & 0.15 & 0.18\\ \hline
 Empirical SD & Debiased SS & 0.30 & 0.23 & 0.21 & 0.22 & 0.18 & 0.24 & 0.26 & 0.29\\
  & MLE SS & 0.58 & 0.41 & 0.36 & 0.38 & 0.31 & 0.40 & 0.48 & 0.56\\
  & Debiased MS & 0.17 & 0.14 & 0.14 & 0.14 & 0.12 & 0.14 & 0.14 & 0.18\\
  & MLE MS & 0.29 & 0.22 & 0.22 & 0.23 & 0.20 & 0.21 & 0.24 & 0.30\\
  & Oracle $(n_2)$ & 0.28 & 0.22 & 0.20 & 0.22 & 0.19 & 0.20 & 0.22 & 0.26\\
  & Oracle $(n)$ & 0.17 & 0.14 & 0.14 & 0.16 & 0.14 & 0.14 & 0.14 & 0.17\\ \hline
 \end{tabular*}}
\bigskip
\end{table}

The simulation results for compound symmetry correlation structure are summarized in Table \ref{table-cs-5}.
The same trends concerning the bias and large variability of the MLE seen in Table \ref{table-ar1-5} are also present in this setting.
Again the debiased lasso estimators are approximately unbiased, and the multiple splitting debiased lasso estimator has approximate 95\% coverage for each coefficient.
Multiple splitting again greatly reduces the variability of single split estimators, resulting in thinner confidence intervals with more power for detecting small coefficient values.

\begin{table}
\caption{Logistic regression simulation results for $n=500$, $p=700$, $s_0=6$, and compound symmetry correlation structure with parameter 0.5. Selection results refer to the lasso in the single split estimator, where the average selected model size was 37. The fitted model for estimating each $\beta_j$ was $\Slass\cup\{j\}$. Nominal confidence interval coverage probabilities are 0.95.} \label{table-cs-5}
{\begin{tabular*}{\textwidth}{l l r r r r r r r r}
\hline  & Estimator & $\beta_{489}$ & $\beta_{130}$ & $\beta_{680}$ & $\beta_{488}$ & $\beta_{476}$ & $\beta_{190}$ & $\beta_{510}$ & $\beta_{336}$\\ \hline
$\bbeta^0_j$ &  & -1.50 & -1.00 & -0.50 & 0.00 & 0.00 & 0.50 & 1.00 & 1.50\\
 Selection Rate &  & 1.00 & 0.96 & 0.51 & 0.07 & 0.05 & 0.44 & 0.98 & 1.00\\ \hline
 Bias & Debiased SS & 0.01 & 0.04 & 0.04 & -0.02 & 0.00 & 0.00 & -0.02 & -0.05\\
  & MLE SS & -0.42 & -0.27 & -0.13 & -0.02 & -0.01 & 0.19 & 0.30 & 0.37\\
  & Debiased MS & 0.02 & 0.03 & 0.03 & -0.01 & 0.00 & -0.02 & -0.02 & -0.02\\
  & MLE MS & -0.39 & -0.26 & -0.14 & -0.01 & 0.00 & 0.15 & 0.26 & 0.39\\
  & Oracle $(n_2)$ & -0.05 & -0.04 & -0.02 & 0.00 & 0.00 & 0.01 & 0.04 & 0.07\\
  & Oracle $(n)$ & -0.03 & -0.02 & -0.02 & -0.01 & 0.00 & 0.02 & 0.02 & 0.03\\ \hline
 Coverage & Debiased SS & 0.88 & 0.92 & 0.92 & 0.94 & 0.96 & 0.92 & 0.90 & 0.89\\
  & MLE SS & 0.80 & 0.89 & 0.91 & 0.90 & 0.93 & 0.86 & 0.85 & 0.82\\
  & Debiased MS & 0.94 & 0.96 & 0.94 & 0.96 & 0.96 & 0.94 & 0.94 & 0.96\\
  & MLE MS & 0.68 & 0.84 & 0.90 & 0.96 & 0.94 & 0.92 & 0.81 & 0.76\\
  & Oracle $(n_2)$ & 0.94 & 0.94 & 0.95 & 0.91 & 0.94 & 0.92 & 0.96 & 0.94\\
  & Oracle $(n)$ & 0.96 & 0.98 & 0.96 & 0.92 & 0.92 & 0.96 & 0.96 & 0.98\\ \hline
 Rejection Rate & Debiased SS & 1.00 & 0.99 & 0.49 & 0.06 & 0.04 & 0.57 & 0.98 & 0.99\\
  ($H_0: \beta_j=0$) & MLE SS & 1.00 & 1.00 & 0.56 & 0.10 & 0.07 & 0.63 & 0.98 & 0.99\\
  & Debiased MS & 1.00 & 1.00 & 0.82 & 0.04 & 0.04 & 0.87 & 1.00 & 1.00\\
  & MLE MS & 1.00 & 1.00 & 0.84 & 0.04 & 0.06 & 0.88 & 1.00 & 1.00\\
  & Oracle $(n_2)$ & 1.00 & 1.00 & 0.66 & 0.09 & 0.06 & 0.66 & 1.00 & 1.00\\
  & Oracle $(n)$ & 1.00 & 1.00 & 0.92 & 0.07 & 0.07 & 0.92 & 1.00 & 1.00\\ \hline
 Standard Error & Debiased SS & 0.25 & 0.23 & 0.23 & 0.23 & 0.23 & 0.23 & 0.23 & 0.25\\
  & MLE SS & 0.36 & 0.32 & 0.30 & 0.28 & 0.29 & 0.30 & 0.32 & 0.36\\
  & Debiased MS & 0.19 & 0.17 & 0.16 & 0.15 & 0.15 & 0.16 & 0.17 & 0.19\\
  & MLE MS & 0.27 & 0.23 & 0.21 & 0.21 & 0.21 & 0.21 & 0.23 & 0.27\\
  & Oracle $(n_2)$ & 0.27 & 0.24 & 0.22 & 0.22 & 0.22 & 0.22 & 0.24 & 0.27\\
  & Oracle $(n)$ & 0.18 & 0.17 & 0.15 & 0.15 & 0.15 & 0.15 & 0.17 & 0.18\\ \hline
 Empirical SD & Debiased SS & 0.29 & 0.25 & 0.26 & 0.22 & 0.24 & 0.26 & 0.27 & 0.29\\
  & MLE SS & 0.56 & 0.43 & 0.39 & 0.34 & 0.38 & 0.40 & 0.50 & 0.53\\
  & Debiased MS & 0.18 & 0.16 & 0.16 & 0.15 & 0.15 & 0.16 & 0.17 & 0.18\\
  & MLE MS & 0.28 & 0.24 & 0.22 & 0.22 & 0.22 & 0.21 & 0.24 & 0.27\\
  & Oracle $(n_2)$ & 0.28 & 0.25 & 0.23 & 0.25 & 0.23 & 0.23 & 0.25 & 0.29\\
  & Oracle $(n)$ & 0.18 & 0.15 & 0.16 & 0.16 & 0.16 & 0.15 & 0.16 & 0.17\\ \hline
 \end{tabular*}}
\bigskip
\end{table}

Lastly, we assess the performance of the post-model selection procedure that does not pre-specify any covariate of interest.
This simulation setting is identical to that of Table \ref{table-ar1-5} with AR(1) correlation structure, but the estimates are now all based on a single model fit on only the selected covariates.
For covariates that are not selected, their coefficient estimate and standard error are both set to zero.
The oracle MLE results we present are estimated on $S_0$ instead of $S_0\cup \{j\}$.

\begin{table}
\caption{Logistic regression simulation results for $n=500$, $p=700$, $s_0=6$, and AR(1) correlation structure with parameter 0.5. Selection results refer to the lasso in the single split estimator, where the average selected model size was 41. The fitted model for estimating each $\beta_j$ was $\Slass$. Nominal confidence interval coverage probabilities are 0.95.} \label{table-postselection-ar1-5}
{\begin{tabular*}{\textwidth}{l l r r r r r r r r}
\hline  & Estimator & $\beta_{489}$ & $\beta_{130}$ & $\beta_{680}$ & $\beta_{488}$ & $\beta_{476}$ & $\beta_{190}$ & $\beta_{510}$ & $\beta_{336}$\\ \hline
$\bbeta^0_j$ &  & -1.50 & -1.00 & -0.50 & 0.00 & 0.00 & 0.50 & 1.00 & 1.50\\
 Selection Rate &  & 1.00 & 1.00 & 0.64 & 0.13 & 0.07 & 0.69 & 1.00 & 1.00\\ \hline
 Bias & Debiased SS & 0.00 & 0.06 & 0.18 & -0.01 & 0.00 & -0.18 & 0.00 & 0.00\\
  & MLE SS & -0.66 & -0.35 & 0.02 & -0.02 & 0.01 & -0.03 & 0.44 & 0.66\\
  & Oracle $(n_2)$ & -0.09 & -0.05 & 0.00 & 0.00 & 0.00 & 0.04 & 0.07 & 0.09\\ \hline
 Coverage & Debiased SS & 0.83 & 0.90 & 0.57 & 0.99 & 1.00 & 0.59 & 0.87 & 0.89\\
  & MLE SS & 0.70 & 0.83 & 0.51 & 0.98 & 1.00 & 0.57 & 0.75 & 0.66\\
  & Oracle $(n_2)$ & 0.95 & 0.96 & 0.94 & 1.00 & 1.00 & 0.94 & 0.96 & 0.97\\ \hline
 Rejection Rate & Debiased SS & 1.00 & 0.99 & 0.47 & 0.01 & 0.00 & 0.42 & 1.00 & 1.00\\
  ($H_0: \beta_j=0$) & MLE SS & 1.00 & 0.99 & 0.49 & 0.02 & 0.00 & 0.45 & 1.00 & 1.00\\
  & Oracle $(n_2)$ & 1.00 & 1.00 & 0.74 & 0.00 & 0.00 & 0.80 & 1.00 & 1.00\\ \hline
 Standard Error & Debiased SS & 0.22 & 0.20 & 0.12 & 0.03 & 0.01 & 0.13 & 0.20 & 0.22\\
  & MLE SS & 0.40 & 0.32 & 0.18 & 0.04 & 0.02 & 0.20 & 0.33 & 0.40\\
  & Oracle $(n_2)$ & 0.26 & 0.22 & 0.20 & 0.00 & 0.00 & 0.20 & 0.22 & 0.26\\ \hline
 Empirical SD & Debiased SS & 0.32 & 0.23 & 0.30 & 0.10 & 0.04 & 0.30 & 0.26 & 0.29\\
  & MLE SS & 0.83 & 0.49 & 0.47 & 0.18 & 0.07 & 0.50 & 0.55 & 0.75\\
  & Oracle $(n_2)$ & 0.28 & 0.22 & 0.20 & 0.00 & 0.00 & 0.20 & 0.22 & 0.26\\ \hline
 \end{tabular*}}
\bigskip
\end{table}

The simulation results are summarized in Table \ref{table-postselection-ar1-5}.
For small coefficients, there is poor confidence interval coverage across all non-oracle estimators due to the randomness associated with their inclusion in each model.
For larger coefficients that are nearly always selected by the lasso, the performance resembles that of Table \ref{table-ar1-5}.
These results demonstrate the importance of only analyzing coefficients that are included in the fitted model.
For larger sample sizes, the lasso selection performance can improve substantially, leading to improved coverage and bias correction of the debiased lasso after model selection. See \ref{appendix:sims} for additional simulations results with a larger sample size. We also provide additional simulation results with higher correlations among covariates in \ref{appendix:sims}.

\section{Real Data Example: The Mid-South Tobacco Case-Control Study}
\label{s:analysis}

We apply the proposed method to a dataset of single nucleotide polymorphisms (SNPs) from a sample of African-American participants in the Mid-South Tobacco Case-Control study population \citep{jiang2019exome, xu2020prediction, han2022identification} to assess genetic risk factors for nicotine dependence.
The dataset is publicly available with GEO accession GSE148375.
It originally contained 242,901 SNPs measured across 3399 individuals.
We excluded SNPs that were insertions or deletions, had a call rate less than 95\%, were not in Hardy-Weinberg equilibrium ($p>10^{-6}$), or had a minor allele frequency of less than 0.01.
Subjects missing more than 1\% of the remaining SNPs were excluded, and missing values were then imputed using the observed allele frequencies for each SNP.
After data cleaning, the covariates consist of 32,557 SNPs as well as gender and age. The response variable is a binary indicator of smoking status (1=smoker), where 1607 of the 3317 participants are smokers.

Prior research has identified several genetic regions with SNPs that are associated with nicotine dependence, including 15q25.1 (CHRNA5/A3/B4), 8p11.21 (CHRNB3/A6), and 19q13.2 (CYP2A6/A7/B6, EGLN2, RAB4B, and NUMBL).
See, for example, \cite{yang2016converging} and the references therein.
We choose the target covariates to be the 16 measured SNPs that lie in these three regions, as well as the demographic variables.

The results from our multiple splitting debiased lasso estimator are presented in Table \ref{table-data-analysis}.
After adjusting for multiple testing using the Holm procedure \citep{holm1979simple}, none of the SNPs have a significant association with smoking status.
The SNP rs3733829 has the largest coefficient estimate, with an estimated 37\% increase in the odds of being a smoker and unadjusted 95\% confidence interval (9\%, 73\%), controlling for all other covariates.
The association of rs3733829 with increased cigarette consumption has been discussed by \cite{tobacco2010genome} and \cite{bloom2014variants}.

\begin{table}
\caption{Multiple splitting debiased lasso estimates for a logistic regression model of smoking status on standardized age, gender, and 32,557 SNPs, based on a case-control sample of 3317 individuals. There are 16 measured SNPs that lie in the regions of interest after data cleaning. P-values are adjusted for multiple testing using the Holm procedure and truncated at one, while confidence intervals are not adjusted.} \label{table-data-analysis}
{\begin{tabular*}{\textwidth}{l r r r r r r r r r r}
\hline
Covariate & Gene & $\hat\beta$ & SE & Holm P-value & Odds Ratio (95\% CI)\\
\hline
Intercept &  & -0.01 & 0.18 & 1.00 & 0.99 (0.70, 1.41)\\
Age &  &  0.03 & 0.03 & 1.00 & 1.03 (0.96, 1.09)\\
Male &  & 0.45 & 0.07 & $1.76\times 10^{-10}$ & 1.58 (1.38, 1.79)\\
rs35327613 & CHRNB3 & -0.06 & 0.11 & 1.00 & 0.94 (0.76, 1.17)\\
rs61740655 & CHRNA5 & -0.10 & 0.13 & 1.00 & 0.90 (0.70, 1.17)\\
rs79109919 & CHRNA5 & -0.06 & 0.16 & 1.00 & 0.94 (0.68, 1.29)\\
rs16969968 & CHRNA5 & -0.09 & 0.11 & 1.00 & 0.91 (0.74, 1.13)\\
rs938682 & CHRNA3 & 0.10 & 0.10 & 1.00 & 1.11 (0.91, 1.35)\\
rs8042374 & CHRNA3 & -0.18 & 0.12 & 1.00 & 0.83 (0.66, 1.05)\\
rs61737502 & CHRNB4 & 0.04 & 0.10 & 1.00 & 1.04 (0.85, 1.28)\\
rs56218866 & CHRNB4 & -0.10 & 0.14 & 1.00 & 0.91 (0.68, 1.21)\\
rs950776 & CHRNB4 & -0.02 & 0.08 & 1.00 & 0.98 (0.83, 1.15)\\
rs12440298 & CHRNB4 & -0.03 & 0.06 & 1.00 & 0.97 (0.87, 1.09)\\
rs3865452 & COQ8B & -0.06 & 0.06 & 1.00 & 0.94 (0.84, 1.04)\\
rs3733829 & EGLN2 & 0.32 & 0.12 & 0.13 & 1.37 (1.09, 1.73)\\
rs75152309 & CYP2A7 & 0.07 & 0.09 & 1.00 & 1.07 (0.90, 1.28)\\
rs73032311 & CYP2A7 & -0.09 & 0.08 & 1.00 & 0.91 (0.78, 1.07)\\
rs28399499 & CYP2B6 & 0.00 & 0.10 & 1.00 & 1.00 (0.82, 1.22)\\
rs7260329 & CYP2B6 & -0.16 & 0.08 & 0.78 & 0.86 (0.73, 1.00)
 \end{tabular*}}
\bigskip
\end{table}

\section*{Acknowledgements}

This work was supported in part by NIH Grants R01AG056764 and RF1AG075107,  and by NSF Grant DMS-1915711. 

\bibliographystyle{apalike} 
\bibliography{references.bib}

\appendix

\clearpage
\renewcommand{\thesection}{Appendix \Alph{section}}

\section{Proof of Theorem 1: Results for GLM lasso selected model}\label{appendix:thm1}

In order to show the model selection properties of the GLM lasso under Assumptions 1-5, we consider the results on false negatives and false positives (i.e. selected model size) separately in Lemmas 1 and 2 below.
These are each based on oracle inequalities from \cite{huang2012estimation}, specifically using the unweighted lasso, target vector $\bbeta^0$, and target set $S = S_0\cup \{1\}$ that includes the intercept.
We use $c_1$ and $c_2$ as generic constants in our probability bounds and denote the sample size as $n$.

\begin{lemma} \label{lemma1}
	Under Assumptions 1-5, if the diagonal entries of $\bIX^T\bIX$ are bounded above and away from zero and the compatibility constant \citep{huang2012estimation} is bounded below, then the GLM lasso has no false negatives with probability at least $1-c_1\exp(-c_2\log(p))$ for constants $c_1,c_2>0$.
\end{lemma}

This lemma is a direct result of Theorem 9 (iii) of \cite{huang2012estimation}, which gives an $\ell_1$ bound for the GLM lasso estimation error.
We proceed by verifying that our assumptions imply the required conditions in Theorem 9 of \cite{huang2012estimation} are satisfied.

We first use the bound $\log x \in [1-1/x, x-1]$, to verify the required Lipschitz condition
\begin{eqnarray*}
&&  \max_{i\leq n} \left |\log A''(\bIx_i^T\bbeta^0 + t) - \log A''(\bIx_i^T\bbeta^0)\right | \\
&& \qquad \qquad  = \max_{i\leq n} \left |\log \frac{A''(\bIx_i^T\bbeta^0 + t)}{A''(\bIx_i^T\bbeta^0)}\right | \\
  && \qquad \qquad \leq \max_{i\leq n}\frac{|A''(\bIx_i^T\bbeta^0 + t)- A''(\bIx_i^T\bbeta^0)|}{\min\left (A''(\bIx_i^T\bbeta^0), A''(\bIx_i^T\bbeta^0 + t)\right )} \\
  && \qquad \qquad \leq \frac{c_{Lip} |t|}{\inf_{i\leq n, |t|\leq \delta} A''(\bIx_i^T\bbeta^0 + t)}
\end{eqnarray*}
for all $|t|\leq \delta$, where the denominator is bounded below since $\norm{\bIX\bbeta^0}{\infty}$ is bounded above almost surely and $A''$ is a positive function.

Next, we can choose penalty levels $\lambda>\lambda_1=\lambda_0$ each of order $\sqrt{\log(p)/n}$ so that (31) of \cite{huang2012estimation} is satisfied for large $n$ when the compatibility constant $\kappa_*(\xi, S)$ is bounded below and $|S|\sqrt{\log(p)/n}=\littleO{1}$, as we have assumed.

Lastly, to verify condition (29) of \cite{huang2012estimation}, note that the positive quantities $A''(\bIx_i^T\bbeta^0), i=1,\ldots,n$ are bounded almost surely, so the diagonal elements of $P_n\ddot{\rho}_{\bbeta^0}= \frac{1}{n}\sum_{i=1}^n A''(\bIx_i^T\bbeta^0)\bIx\bIx_i^T$ are bounded above and away from zero due to the assumed corresponding bounds on $\frac{1}{n}\sum_{i=1}^n x_{ij}^2$ for all $j=1,\ldots,p+1$.

Now that we have verified the conditions, by Theorem 9 of \cite{huang2012estimation} using the seminorm $\phi(\bb) = \norm{\bb_S}{1}/|S|$, with probability at least $1-c_1\exp(-c_2\log(p))$ for constants $c_1,c_2>0$, we have $\norm{\bbetal_S - \bbeta_S^0}{1}\leq k|S|\sqrt{\log(p)/n}$ for some constant $k>0$.
Therefore, on this event, there are no false negatives if  $\min_{j\in S}|\bbeta_j^0| > k(s_0+1) \sqrt{\log(p)/n}$, as we have assumed.

\begin{lemma} \label{lemma2}
Under the conditions of Lemma 1, the GLM lasso selected model size has an upper bound proportional to the true model size $s_0$, with probability at least $1-c_1\exp(c_2n)$ for some constants $c_1,c_2>0$.
\end{lemma}

This lemma is a direct result of Theorem 19 of \cite{huang2012estimation}, which gives an upper bound for the number of false positives selected by the GLM lasso in terms of a compatibility constant and restricted upper eigenvalue.

We begin by showing that the restricted upper eigenvalue 
\[
  \kappa_+(s) = \sup_{|B|=s, B\cap S=\emptyset, \bb = \in\Cone{\xi}{S}, M_3\norm{\bb_S}{1}\leq \eta^*}
  \lambda_{max}\left (\int_0^1 \frac{1}{n}\sum_{i=1}^n A''(\bIx_i^T\bbeta^0 + t\bIx_i^T\bb)\left [\bIx_i\bIx_i^T\right ]_B dt\right )
\]
over the cone $\Cone{\xi}{S} = \{\bb\in R^{p+1}: \norm{\bb_{S^c}}{1}\leq \xi\norm{\bb_S}{1}\neq 0\}$ is bounded above with probability tending to one when $s$ is proportional to $s_0$.
For large $n$ and small $\eta^*$, the term involving $A''$ is bounded almost surely by our Lipschitz assumptions since $\norm{\bb}{1} = \norm{\bb_S}{1} + \norm{\bb_{S^c}}{1}\leq (1+\xi)\norm{\bb_S}{1}$ so $|\bIx_i^T\bb|\leq \norm{\bIx_i}{\infty}\norm{\bb}{1}\leq K(1+\xi)\eta^*$ in the cone $\Cone{\xi}{S}$.
Since the intercept is included in $S$, and the eigenvalues of $E[\bIx\bIx^T]$ are bounded above, we can apply Lemma 15 of \cite{loh2012sparseeigen} for zero-mean subgaussian random vectors to show that $\sup_{|B|\leq s, B\cap S=\emptyset}\lambda_{\max}\left ( \frac{1}{n}\left [\bIX^T\bIX\right ]_B\right )^2$ is bounded above by a constant with probability at least $1-c_1\exp(c_2n)$ for some constants $c_1,c_2>0$ when $s\propto s_0$ and $s_0\log(p)/n = \littleO{1}$. 
Using this upper bound for the upper restricted eigenvalue and the assumed lower bound for the compatibility constant, Theorem 19 of \cite{huang2012estimation} implies that the number of false positives, and thus the selected model size, is at most of order $s_0$.

\bigskip

We now are ready to prove Theorem 1. We begin by providing probability bounds for the events assumed in Lemmas 1 and 2. 
First, \cite{van2009conditions} showed that the compatibility constant is bounded away from zero with probability at least $1-c_1\exp(-c_2n)$ for some constants $c_1,c_2>0$.

Next, we seek to bound $\frac{1}{n}\sum_{i=1}^n x_{ij}^2$ from above and away from zero uniformly over $j=1,\ldots,p+1$.
This is trivial for the intercept, and the rest of the covariates are each bounded almost surely with mean zero and a variance of at least $min_j \bSig_{jj}\geq \lambda_{min}(\bSig)>0$. We apply Hoeffding's inequality and a union bound to get
\begin{eqnarray*}
  P\left (\max_{2\leq j\leq p+1}\left |\frac{1}{n}\sum_{i=1}^n x_{ij}^2 - \bSig_{jj}\right | \geq t\right )
  &\leq& \sum_{j=2}^{p+1} P\left (\left |\frac{1}{n}\sum_{i=1}^n x_{ij}^2 - \bSig_{jj}\right |\geq t\right ) \\
  &\leq&  \sum_{j=2}^{p+1} 2\exp\left (-\frac{2nt^2}{K^4} \right ) \\
  &\leq& \exp\left (-\frac{n\left [2t^2 - K^4\log(2p)/n\right ]}{K^4} \right ).
\end{eqnarray*}
Since $\log(p)/n=\op{1}$, for large $n$ we can fix a $t\in (0, \lambda_{min}(\bSig))$ and the probability bound will be of order $c_1\exp(-c_2n)$ for some positive constants $c_1$ and $c_2$.

Lastly, let $\mathcal{B}$ denote the union of events that the compatibility constant is bounded, the diagonal entries of $\bIX^T\bIX$ are bounded, and the upper restricted eigenvalue (see Lemma 2) is bounded.
Since the probability bound on each of these events has the same order, by a union bound and rescaling of the constants involved we have $P(\mathcal{B}^c)\leq c_1\exp(-c_2n)$ for some $c_1,c_2>0$.
Then for the event $\mathcal{A} = \{\Slass\supseteq S_0, |\Slass|\leq ks_0\}$ with $k>0$ not depending on $n$ and some constants $c_3,c_4,c_5,c_6>0$ we have
\begin{eqnarray*}
  P(\mathcal{A})
  &=& P(\mathcal{A}|\mathcal{B})P(\mathcal{B}) + P(\mathcal{A}|\mathcal{B}^c)P(\mathcal{B}^c) \\
  &\geq& \underbrace{\left [1-c_3\exp(-c_4\log(p))\right ]}_{by\ Lemma\ 1}\left [1-c_1\exp(-c_2n)\right ] + 0 \\
  &\geq& 1 - c_5\exp(-c_6\log(p))
\end{eqnarray*}
for $n$ sufficiently large, since $\log(p)/n = \littleO{1}$.

\section{Proof of Theorem 2: Asymptotic results for debiased lasso after single sample split}\label{appendix:thm2}

Suppose the sample of $n$ individuals is split into a subsample $\mathcal{D}_1$ of size $n_1=qn$ and a subsample $\mathcal{D}_2$ of size $n_2 = (1-q)n$, where we assume $n_1$ is an integer and $q\in (0,1)$. 
We apply a model selection procedure to $\mathcal{D}_1$ such that the selected model $\Slass$ satisfies the event $\mathcal{A} = \{\Slass\supseteq S_0, |\Slass|\leq \littleO{\sqrt{n}} \}$ with probability tending to one as $n\rightarrow\infty$, such as the lasso.
For the selected model $\Slass$, which includes the pre-specified set $S$, let $\bbetadl_{\Slass}$ denote the debiased lasso estimator of $\bbeta^0_{\Slass}$ based on the subsample $\mathcal{D}_2$, and similarly denote the asymptotic covariance estimator by $\boldsymbol{\widehat\Theta}_{\Slass}$.
We decompose
\[
   \sqrt{n_2}\ba_{\Slass}^T\left (\bbetadl_{\Slass} - \bbeta^0_{\Slass}\right )
   = -\frac{1}{\sqrt{n_2}}\ba_{\Slass}^T\bSig_{\bbeta^0, \Slass}^{-1}\sum_{i\in \mathcal{D}_2}\dot{\rho}_{\bbeta_{\Slass}^0}(y_i, (\bIx_i)_{\Slass}) + r_{n}
\]
where, conditional on any sequence of selected models $\Slass$ independent of $\mathcal{D}_2$ and satisfying the event $\mathcal{A}$, $r_n\rightarrow_p 0$ by the theoretical results of \cite{xia2021debiased}. 
Hence for any $\varepsilon > 0$,
\begin{eqnarray*}
    P\left (|r_n| \geq \varepsilon\right )
    &=& P\left (|r_n| \geq \varepsilon|\mathcal{A}\right )P(\mathcal{A}) + P\left (|r_n| \geq \varepsilon|\mathcal{A}^c\right )P(\mathcal{A}^c) \\
    &=& P\left (|r_n| \geq \varepsilon|\mathcal{A}\right )(1-\littleO{1}) + \bigO{1}\littleO{1},
\end{eqnarray*}
where $P\left (|r_n|\geq \varepsilon|\mathcal{A}\right )\rightarrow 0$ as $n\rightarrow\infty$ by \cite{xia2021debiased} as mentioned above, so $r_n = \op{1}$ unconditionally as $n\rightarrow\infty$.
Asymptotic normality, conditional on $\mathcal{A}$, also follows from the low-dimensional results of \cite{xia2021debiased} using the Lindeberg-Feller Central Limit Theorem.

\section{Proof of Theorem 3: Asymptotic results for debiased lasso after multiple sample splits}\label{appendix:thm3}

First we apply the results from the proof of Theorem 2 to the $(p+1)$-vector $\ba$ with $\norm{\ba_S}{2}=\norm{\ba}{2}=1$, i.e. a contrast of only the pre-specified covariates $S$. 
The single split debiased lasso estimator $\bbetadl_b$ satisfies
\begin{eqnarray*} 
  \sqrt{n_2}\ba_{S}^T(\bbetadl_{S,b} - \bbeta_S^0) 
  &=& -\frac{1}{\sqrt{n_2}}\ba_{\Slass_b}^T\bSig_{\bbeta^0, \Slass_b}^{-1}
  \sum_{i\in D_{2,b}}\dot{\rho}_{\bbeta_{\Slass_b}^0}(y_i, (\bIx_i)_{\Slass_b}) + \op{1} \\
  &=& \frac{1}{\sqrt{n_2}}\ba_{\Slass_b}^T\bSig_{\bbeta^0, \Slass_b}^{-1}\Isub_{\Slass_b}  \sum_{i=1}^n v_{b,i}\dot{\rho}_{\bbeta^0}(y_i, \bIx_i) + \op{1},
\end{eqnarray*}
where $\Slass_b\supseteq S_0\cup S$ is the fitted model, $\Isub_{\Slass_b}$ is the $|\Slass_b|\times (p+1)$ matrix such that $\Isub_{\Slass_b}\bb = \bb_{\Slass_b}$ for any $(p+1)$-vector $\bb$, and $v_{b,i}$ is the indicator variable equal to one for $i\in D_{2,b}$.

Now consider the multiple splitting estimator $\bbetams$ obtained by averaging across all possible splits $(D_{1,b}, D_{2,b})_{b=1}^B$, $B = {n\choose n_2}$.
Again letting $\Slass_{b}$ denote the fitted model from data $\mathcal{Z}=(y_i,\bIx_i)_{i=1}^n$ split by sampling indicators $\bv_b$, the multiple splitting estimator satisfies
\begin{eqnarray}\label{ms-decomp}
  \sqrt{n}\ba_S^T(\bbetams_S - \bbeta_S^0)
  &=& \frac{(1-q)^{-1}}{\sqrt{n}}\sum_{i= 1}^n E\left [\ba_{\Slass_b}^T\bSig_{\bbeta^0, \Slass_b}^{-1}\Isub_{\Slass_b}   v_{b,i} | \mathcal{Z}\right ]\dot{\rho}_{\bbeta^0}(y_i, \bIx_i) + \op{1} \\
  &=& \frac{1}{\sqrt{n}}\sum_{i= 1}^n\left (y_i - A'(\bIx_i^T\bbeta^0) \right )\etavec^T\bIx_i \nonumber \\
  &&+ \underbrace{\frac{1}{\sqrt{n}}\sum_{i= 1}^n \left ( (1-q)^{-1}E\left [\ba_{\Slass_b}^T\bSig_{\bbeta^0, \Slass_b}^{-1}\Isub_{\Slass_b}v_{b,i} | \mathcal{Z}\right ] - \etavec^T\right )\dot{\rho}_{\bbeta^0}(y_i, \bIx_i)}_{r_n} + \op{1}, \nonumber
\end{eqnarray} 
where the expectations are with respect to the random splits $b=1,\ldots,B$, conditional on the data.

To establish asymptotic normality, we first show that $r_n = \op{1}$.
Let $\boldsymbol{\widetilde{v}}_b$ be another sampling indicator vector independent of $\bv_b$, with fitted model $\widetilde{S}_b$.
The expectations in the following decomposition are taken with respect to $\bv_b$ and $\boldsymbol{\tilde{v}}_b$, i.e. over $b=1,\ldots,B$, conditional on the data:
\begin{eqnarray*}
 && (1-q)^{-1}E\left [\ba_{\Slass_b}^T\bSig_{\bbeta^0, \Slass_b}^{-1}\Isub_{\Slass_b}v_{b,i} | \mathcal{Z}\right ] - \etavec^T \\
 && \qquad
  = (1-q)^{-1}E\left [\ba_{\widetilde{S}_b}^T\bSig_{\bbeta^0, \widetilde{S}_b}^{-1}\Isub_{\widetilde{S}_b}v_{b,i} | \mathcal{Z}\right ] - \etavec^T \\
&& \qquad \qquad  + \ (1-q)^{-1}\left (E\left [\ba_{\Slass_b}^T\bSig_{\bbeta^0, \Slass_b}^{-1}\Isub_{\Slass_b}v_{b,i} | \mathcal{Z}\right ] 
  - E\left [\ba_{\widetilde{S}_b}^T\bSig_{\bbeta^0, \widetilde{S}_b}^{-1}\Isub_{\widetilde{S}_b}v_{b,i} | \mathcal{Z}\right ]\right ),
\end{eqnarray*}
and, by independence, $E\left [\ba_{\widetilde{S}_b}^T\bSig_{\bbeta^0, \widetilde{S}_b}^{-1}\Isub_{\widetilde{S}_b}v_{b,i} | \mathcal{Z}\right ] = E\left [\ba_{\widetilde{S}_b}^T\bSig_{\bbeta^0, \widetilde{S}_b}^{-1}\Isub_{\widetilde{S}_b}|\mathcal{Z}\right ]P(v_{b,i}=1)$ with $P(v_{b,i}=1)=n_2/n = 1-q$.
Since this quantity does not depend on $i$, we can write
\[
  r_n
  = \underbrace{\left (E\left [\ba_{\widetilde{S}}^T\bSig_{\bbeta^0, \widetilde{S}}^{-1}\Isub_{\widetilde{S}} | \mathcal{Z}\right ] - \etavec^T\right )\sqrt{n} P_n \dot{\rho}_{\bbeta^0}}_{r_{n,1}} 
  + \underbrace{\frac{(1-q)^{-1}}{\sqrt{n}}\sum_{i= 1}^n h_{i,n}}_{r_{n,2}}
\]
with $h_{i,n} 
    = \left [E\left (\ba_{\Slass_b}^T\bSig_{\bbeta^0, \Slass_b}^{-1}\Isub_{\Slass_b}v_{b,i}\Big |\mathcal{Z} \right )
    - E\left (\ba_{\widetilde{S}_b}^T\bSig_{\bbeta^0, \widetilde{S}_b}^{-1}\Isub_{\widetilde{S}_b}v_{b,i}\Big |\mathcal{Z} \right )\right ]\dot{\rho}_{\bbeta^0}(y_i,\bIx_i)$
    so $r_{n,2} = \op{1}$ by Assumption (6).
Also, since $P_n\dot{\rho}_{\bbeta^0}$ has bounded entries with mean zero, we apply Hoeffding's inequality and a union bound to obtain
\begin{eqnarray*}
  P\left (\max_{j} \left | \sqrt{n}\left (P_n\dot{\rho}_{\bbeta^0}\right )_j\right | \geq t\sqrt{\log(p)} \right )
  &\leq& p P\left ( \left |\sqrt{n}\left (P_n\dot{\rho}_{\bbeta^0}\right )_j\right | 
  \geq t\sqrt{\log(p)} \right ) \\
  &\leq& 2p\exp\left (-ct^2\log(p)\right )
\end{eqnarray*}
for some constant $c>0$ and $t>0$, so $\norm{\sqrt{n}P_n\dot{\rho}_{\bbeta^0}}{\infty} = \Op{\sqrt{\log(p)}}$ and, applying Assumption (7), we have
\begin{eqnarray*}
  |r_{n,1}|
  &\leq& \norm{E\left [\ba_{\widetilde{S}_b}^T\bSig_{\bbeta^0, \widetilde{S}_b}^{-1}\Isub_{\widetilde{S}_b} | \mathcal{Z}\right ] - \etavec^T}{1}
  \norm{\sqrt{n} P_n \dot{\rho}_{\bbeta^0}}{\infty} \\
  &=& \op{1/\sqrt{\log(p)}} \Op{\sqrt{\log(p)}} = \op{1}.
\end{eqnarray*}

In summary, 
\[
  \sqrt{n}\ba_S^T(\bbetams_S - \bbeta_S^0)
  = \sqrt{n}\etavec^T P_n\dot{\rho}_{\bbeta^0} + \op{1}
  = \frac{1}{\sqrt{n}}\sum_{i= 1}^n\left (y_i - A'(\bIx_i^T\bbeta^0) \right )\etavec^T\bIx_i + o_p(1),
\]
so we have asymptotic normality by the Lindeberg-Feller Central Limit Theorem.
Note that 
\begin{eqnarray*}
    \norm{\etavec}{1}
    \rightarrow_p \norm{E\left [\ba_{\widetilde{S}_b}^T\bSig_{\bbeta^0, \widetilde{S}_b}^{-1}\Isub_{\widetilde{S}_b} | \mathcal{Z}\right ]}{1}
    &\leq& E\left [\norm{\ba_{\widetilde{S}_b}^T\bSig_{\bbeta^0, \widetilde{S}_b}^{-1}}{1} | \mathcal{Z}\right ] \\
    &\leq& E\left [\sqrt{|\widehat{S}_b|}\norm{\ba_{\widetilde{S}_b}^T\bSig_{\bbeta^0, \widetilde{S}_b}^{-1}}{2} | \mathcal{Z}\right ] \\
    &=& \Op{\sqrt{s_0}}
\end{eqnarray*}
since $\ba$ is a unit vector, $\bSig^{-1}_{\bbeta^0}$ has bounded eigenvalues, and $|\Slass_b|\leq ks_0$ with probability tending to one.
Thus the Lindeberg condition can be verified by standard arguments for bounded covariates.

\section{Further discussion of the multiple splitting assumptions}\label{appendix:disc}

The asymptotic representation used in our multiple splitting proof is distinct from the one used by \cite{fei2021estimation} who, in their theoretical arguments, applied a different decomposition of $\sqrt{n}\ba_S^T(\bbetams_S - \bbeta_S^0)$ based on comparing the single split estimators with their respective oracle estimators.
Under their partial orthogonality condition, which requires that the signal variables are independent of the noise variables and thus that $\bSig_{\bbeta^0}$ is block diagonal (reordering the covariates without loss of generality), they argue that $\bbetams_S$ has the same asymptotic normal distribution as the oracle estimator fit on $S\cup S_0$.
Under different assumptions than \cite{fei2021estimation} we also showed that $\sqrt{n}\ba_S^T(\bbetams_S - \bbeta_S^0)$ is asymptotically normal, but with variance $\etavec^T\bSig_{\bbeta^0}\etavec$ that is possibly larger than the oracle variance $\ba_{S\cup S_0}^T\bSig_{\bbeta^0,S\cup S_0}^{-1}\ba_{S\cup S_0}$.
Our assumptions instead concern the limiting behavior of the average (across all random splits that exclude $(y_i,\bIx_i)$ from model selection, i.e. such that $v_{b,i}=1$) $E\left [\ba_{\Slass_b}^T\bSig_{\bbeta^0, \Slass_b}^{-1}\Isub_{\Slass_b}   v_{b,i} | \mathcal{Z}\right ]$, for $i=1,\ldots,n$.
Note that, for each $i$, the corresponding average is independent of $(y_i,\bIx_i)$.
Since the samples are i.i.d., the effect of always excluding the $i$th observation specifically from model selection should be asymptotically negligible (Assumption 6), and furthermore each average should converge in probability to a bounded random vector $\etavec^T$, which is independent of the data, with moderate rate $\norm{E\left [\ba_{\Slass_b}^T\bSig_{\bbeta^0, \Slass_b}^{-1}\Isub_{\Slass_b} | \mathcal{Z}\right ] - \etavec^T}{1} = \op{1/\sqrt{\log(p)}}$ (Assumption 7).  

We can achieve the oracle-level result of \cite{fei2021estimation} under either of two stronger assumptions that also imply our convergence conditions in Assumptions (6) and (7).
First, under model selection consistency we have $\Slass_b=S\cup S_0$ for all random splits, so $\ba_{\Slass_b}^T\bSig_{\bbeta^0, \Slass_b}^{-1}\Isub_{\Slass_b} = \ba_{S\cup S_0}^T\bSig_{\bbeta^0, S\cup S_0}^{-1}\Isub_{S\cup S_0}$, a deterministic vector, for all $b=1,\ldots,B$ and we can directly set
$\etavec^T 
= E\left [\ba_{\Slass_b}^T\bSig_{\bbeta^0, \Slass_b}^{-1}\Isub_{\Slass_b} | \mathcal{Z}\right ]$.
By (\ref{ms-decomp}) we get $\sqrt{n}\ba_S^T(\bbetams_S - \bbeta_S^0) = \sqrt{n}\etavec^T P_n\dot{\rho}_{\bbeta^0} + \op{1}$. Then, using the fact that $\Isub_{S \cup S_0}\bSig_{\bbeta^0}\Isub_{S \cup S_0}^T=\bSig_{\bbeta^0,S\cup S_0}$, the asymptotic variance becomes $\etavec^T\bSig_{\bbeta^0}\etavec 
= \ba_{S\cup S_0}^T\bSig_{\bbeta^0,S\cup S_0}^{-1}\ba_{S\cup S_0}$.
Alternatively, if the covariates $S\cup S_0$ are independent of $(S\cup S_0)^C$ then $\bSig_{\bbeta^0}=E[A''(\bIx_i^T\bbeta^0)\bIx_i\bIx_i^T]$ has a block diagonal structure that also implies 
$\ba_{\Slass_b}^T\bSig_{\bbeta^0, \Slass_b}^{-1}\Isub_{\Slass_b} 
= \ba_{S\cup S_0}^T\bSig_{\bbeta^0, S\cup S_0}^{-1}\Isub_{S\cup S_0}$ for all $b=1,\ldots,B$, using the fact that $\ba_{(S\cup S_0)^C}$ is a zero-vector.
Therefore this independence assumption also implies oracle-level variance.
Note that, in general, this independence assumption is difficult to justify in practice when the true signal variables are unknown.
An exception is in linear models, where it is sufficient to assume the pre-specified covariates $S$ are independent of $S^C$, e.g. by randomized treatment assignment, since $A''(\cdot)$ is a constant.
In this sense our Assumptions (6) and (7), adapted from \cite{wang2020debiased} for linear models, are less restrictive than those used by others to justify multiple splitting procedures such as \cite{fei2021estimation} for GLMs and \cite{fei2019drawing} for linear models.

\section{Additional simulation results}\label{appendix:sims}

Table \ref{table-postselection-ar1-5-n2000} summarizes results from the same simulation setting as the post-model selection procedure of Table 3, but with a larger sample size of $n=2000$, and $p=2800$ covariates.
We see that the selection rate for small coefficients improves, resulting in lower bias and better confidence interval coverage, and the debiased lasso estimator still has a substantial advantage over the MLE.

\begin{table}
\caption{Logistic regression simulation results for $n=2000$, $p=2800$, $s_0=6$, and AR(1) correlation structure with parameter 0.5. Selection results refer to the lasso in the single split estimator, where the average selected model size was 78. The fitted model for estimating each $\beta_j$ was $\Slass$. Nominal confidence interval coverage probabilities are 0.95.} \label{table-postselection-ar1-5-n2000}
{\begin{tabular*}{\textwidth}{l l r r r r r r r r}
\hline  & Estimator & $\beta_{489}$ & $\beta_{130}$ & $\beta_{680}$ & $\beta_{488}$ & $\beta_{137}$ & $\beta_{190}$ & $\beta_{510}$ & $\beta_{336}$\\ \hline
$\bbeta^0_j$ &  & -1.50 & -1.00 & -0.50 & 0.00 & 0.00 & 0.50 & 1.00 & 1.50\\
 Selection Rate &  & 1.00 & 1.00 & 1.00 & 0.11 & 0.03 & 1.00 & 1.00 & 1.00\\ \hline
 Bias & Debiased SS & -0.01 & -0.01 & 0.01 & -0.00 & 0.00 & -0.01 & 0.02 & 0.01\\
  & MLE SS & -0.24 & -0.16 & -0.08 & -0.00 & 0.00 & 0.07 & 0.17 & 0.24\\
  & Oracle $(n_2)$ & -0.03 & -0.02 & -0.00 & 0.00 & 0.00 & 0.01 & 0.03 & 0.03\\ \hline
 Coverage & Debiased SS & 0.89 & 0.90 & 0.98 & 0.99 & 1.00 & 0.91 & 0.93 & 0.92\\
  & MLE SS & 0.64 & 0.71 & 0.90 & 0.99 & 1.00 & 0.86 & 0.74 & 0.65\\
  & Oracle $(n_2)$ & 0.96 & 0.93 & 0.97 & 1.00 & 1.00 & 0.94 & 0.95 & 0.95\\ \hline
 Rejection Rate & Debiased SS & 1.00 & 1.00 & 1.00 & 0.01 & 0.00 & 1.00 & 1.00 & 1.00\\
  ($H_0: \beta_j=0$) & MLE SS & 1.00 & 1.00 & 1.00 & 0.01 & 0.00 & 1.00 & 1.00 & 1.00\\
  & Oracle $(n_2)$ & 1.00 & 1.00 & 1.00 & 0.00 & 0.00 & 1.00 & 1.00 & 1.00\\ \hline
 Standard Error & Debiased SS & 0.11 & 0.10 & 0.09 & 0.01 & 0.00 & 0.09 & 0.10 & 0.11\\
  & MLE SS & 0.15 & 0.13 & 0.11 & 0.01 & 0.00 & 0.11 & 0.13 & 0.15\\
  & Oracle $(n_2)$ & 0.12 & 0.11 & 0.09 & 0.00 & 0.00 & 0.09 & 0.11 & 0.12\\ \hline
 Empirical SD & Debiased SS & 0.14 & 0.11 & 0.10 & 0.03 & 0.01 & 0.11 & 0.11 & 0.13\\
  & MLE SS & 0.21 & 0.17 & 0.12 & 0.05 & 0.02 & 0.14 & 0.16 & 0.22\\
  & Oracle $(n_2)$ & 0.12 & 0.11 & 0.09 & 0.00 & 0.00 & 0.10 & 0.10 & 0.12\\ \hline
 \end{tabular*}}
\bigskip
\end{table}

We also present additional simulation results that have stronger correlation between covariates. 
Table \ref{table-ar1-8} summarizes results from an AR(1) correlation structure with parameter 0.8, and Table \ref{table-cs-8} summarizes results from a compound symmetry correlation structure with parameter 0.8. 
We see that the multiple splitting debiased lasso estimator is fairly robust to high correlations between covariates, even as the model selection performance deteriorates.
The debiased lasso continues to outperform the MLE, having lower bias and variance as well as better confidence interval coverage.

\begin{table}
\caption{Logistic regression simulation results for $n=500$, $p=700$, $s_0=6$, and AR(1) correlation structure with parameter 0.8. Selection results refer to the lasso in the single split estimator, where the average selected model size was 38. The fitted model for estimating each $\beta_j$ was $\Slass\cup\{j\}$. Nominal confidence interval coverage probabilities are 0.95.} \label{table-ar1-8}
{\begin{tabular*}{\columnwidth}{l l r r r r r r r r}
\hline  & Estimator & $\beta_{489}$ & $\beta_{130}$ & $\beta_{680}$ & $\beta_{488}$ & $\beta_{476}$ & $\beta_{190}$ & $\beta_{510}$ & $\beta_{336}$\\ \hline
$\bbeta^0_j$ &  & -1.50 & -1.00 & -0.50 & 0.00 & 0.00 & 0.50 & 1.00 & 1.50\\
 Selection Rate &  & 1.00 & 0.97 & 0.56 & 0.24 & 0.05 & 0.64 & 0.98 & 1.00\\ \hline
 Bias & Debiased SS & 0.04 & 0.03 & 0.00 & -0.01 & -0.03 & -0.02 & -0.02 & -0.02\\
  & MLE SS & -0.54 & -0.37 & -0.24 & 0.02 & -0.05 & 0.21 & 0.39 & 0.55\\
  & Debiased MS & 0.01 & 0.03 & 0.01 & -0.02 & -0.01 & -0.01 & -0.01 & -0.01\\
  & MLE MS & -0.54 & -0.35 & -0.23 & -0.01 & -0.01 & 0.22 & 0.39 & 0.55\\
  & Oracle $(n_2)$ & -0.08 & -0.03 & -0.02 & -0.03 & -0.01 & 0.02 & 0.05 & 0.08\\
  & Oracle $(n)$ & -0.04 & -0.02 & -0.02 & -0.01 & -0.01 & 0.01 & 0.03 & 0.04\\ \hline
 Coverage & Debiased SS & 0.89 & 0.93 & 0.92 & 0.96 & 0.96 & 0.90 & 0.91 & 0.90\\
  & MLE SS & 0.77 & 0.82 & 0.85 & 0.93 & 0.93 & 0.88 & 0.81 & 0.74\\
  & Debiased MS & 0.93 & 0.92 & 0.93 & 0.96 & 0.95 & 0.94 & 0.90 & 0.92\\
  & MLE MS & 0.58 & 0.73 & 0.86 & 0.97 & 0.94 & 0.85 & 0.70 & 0.53\\
  & Oracle $(n_2)$ & 0.93 & 0.96 & 0.95 & 0.93 & 0.94 & 0.96 & 0.94 & 0.95\\
  & Oracle $(n)$ & 0.94 & 0.96 & 0.92 & 0.96 & 0.96 & 0.94 & 0.93 & 0.96\\ \hline
 Rejection Rate & Debiased SS & 0.99 & 0.92 & 0.59 & 0.04 & 0.04 & 0.57 & 0.95 & 1.00\\
 ($H_0: \beta_j=0$) & MLE SS & 0.99 & 0.90 & 0.64 & 0.07 & 0.07 & 0.60 & 0.94 & 0.99\\
  & Debiased MS & 1.00 & 0.99 & 0.90 & 0.04 & 0.05 & 0.88 & 1.00 & 1.00\\
  & MLE MS & 1.00 & 0.99 & 0.90 & 0.03 & 0.06 & 0.89 & 1.00 & 1.00\\
  & Oracle $(n_2)$ & 1.00 & 1.00 & 0.74 & 0.07 & 0.06 & 0.78 & 1.00 & 1.00\\
  & Oracle $(n)$ & 1.00 & 1.00 & 0.98 & 0.04 & 0.04 & 0.98 & 1.00 & 1.00\\ \hline
 Standard Error & Debiased SS & 0.27 & 0.26 & 0.24 & 0.32 & 0.20 & 0.23 & 0.25 & 0.27\\
  & MLE SS & 0.43 & 0.39 & 0.33 & 0.44 & 0.27 & 0.33 & 0.38 & 0.44\\
  & Debiased MS & 0.19 & 0.17 & 0.15 & 0.20 & 0.12 & 0.15 & 0.17 & 0.19\\
  & MLE MS & 0.30 & 0.26 & 0.22 & 0.32 & 0.19 & 0.22 & 0.26 & 0.30\\
  & Oracle $(n_2)$ & 0.26 & 0.22 & 0.20 & 0.32 & 0.19 & 0.20 & 0.22 & 0.25\\
  & Oracle $(n)$ & 0.17 & 0.15 & 0.14 & 0.22 & 0.13 & 0.14 & 0.15 & 0.17\\ \hline
 Empirical SD & Debiased SS & 0.33 & 0.30 & 0.26 & 0.30 & 0.18 & 0.26 & 0.29 & 0.33\\
  & MLE SS & 0.72 & 0.59 & 0.43 & 0.55 & 0.31 & 0.44 & 0.52 & 0.69\\
  & Debiased MS & 0.21 & 0.20 & 0.17 & 0.21 & 0.12 & 0.17 & 0.19 & 0.20\\
  & MLE MS & 0.35 & 0.32 & 0.26 & 0.33 & 0.19 & 0.26 & 0.30 & 0.34\\
  & Oracle $(n_2)$ & 0.29 & 0.22 & 0.20 & 0.34 & 0.19 & 0.20 & 0.23 & 0.26\\
  & Oracle $(n)$ & 0.19 & 0.15 & 0.14 & 0.23 & 0.13 & 0.15 & 0.17 & 0.18\\ \hline
 \end{tabular*}}
\bigskip
\end{table}

\begin{table}
\caption{Logistic regression simulation results for $n=500$, $p=700$, $s_0=6$, and compound symmetry correlation structure with parameter 0.8. Selection results refer to the lasso in the single split estimator, where the average selected model size was 23. The fitted model for estimating each $\beta_j$ was $\Slass\cup\{j\}$. Nominal confidence interval coverage probabilities are 0.95.} \label{table-cs-8}
{\begin{tabular*}{\columnwidth}{l l r r r r r r r r}
\hline  & Estimator & $\beta_{489}$ & $\beta_{130}$ & $\beta_{680}$ & $\beta_{488}$ & $\beta_{476}$ & $\beta_{190}$ & $\beta_{510}$ & $\beta_{336}$\\  \hline
 $\bbeta^0_j$ &  & -1.50 & -1.00 & -0.50 & 0.00 & 0.00 & 0.50 & 1.00 & 1.50\\
 Selection Rate &  & 0.98 & 0.62 & 0.24 & 0.01 & 0.03 & 0.22 & 0.68 & 0.96\\ \hline
 Bias & Debiased SS & 0.04 & 0.04 & 0.03 & -0.02 & 0.00 & -0.05 & -0.08 & -0.06\\
  & MLE SS & -0.14 & -0.09 & -0.04 & -0.02 & 0.01 & 0.01 & 0.04 & 0.11\\
  & Debiased MS & 0.03 & 0.08 & 0.02 & -0.02 & 0.00 & -0.02 & -0.07 & -0.07\\
  & MLE MS & -0.17 & -0.07 & -0.07 & -0.03 & -0.01 & 0.06 & 0.08 & 0.13\\
  & Oracle $(n_2)$ & -0.07 & 0.03 & -0.07 & -0.03 & 0.00 & 0.05 & 0.03 & 0.02\\
  & Oracle $(n)$ & -0.03 & 0.02 & -0.03 & -0.01 & 0.00 & 0.03 & 0.01 & 0.01\\ \hline
 Coverage & Debiased SS & 0.90 & 0.91 & 0.91 & 0.91 & 0.90 & 0.94 & 0.91 & 0.91\\
  & MLE SS & 0.89 & 0.89 & 0.91 & 0.91 & 0.89 & 0.93 & 0.88 & 0.90\\
  & Debiased MS & 0.90 & 0.94 & 0.91 & 0.96 & 0.96 & 0.93 & 0.92 & 0.90\\
  & MLE MS & 0.86 & 0.90 & 0.93 & 0.96 & 0.95 & 0.93 & 0.88 & 0.88\\
  & Oracle $(n_2)$ & 0.94 & 0.94 & 0.94 & 0.96 & 0.96 & 0.94 & 0.96 & 0.96\\
  & Oracle $(n)$ & 0.94 & 0.94 & 0.92 & 0.96 & 0.93 & 0.94 & 0.92 & 0.94\\ \hline
 Rejection Rate & Debiased SS & 1.00 & 0.82 & 0.33 & 0.09 & 0.10 & 0.33 & 0.77 & 0.99\\
  ($H_0: \beta_j=0$) & MLE SS & 1.00 & 0.85 & 0.36 & 0.09 & 0.11 & 0.34 & 0.78 & 0.98\\
  & Debiased MS & 1.00 & 0.98 & 0.59 & 0.04 & 0.04 & 0.56 & 0.98 & 1.00\\
  & MLE MS & 1.00 & 0.98 & 0.58 & 0.04 & 0.05 & 0.58 & 0.98 & 1.00\\
  & Oracle $(n_2)$ & 1.00 & 0.88 & 0.47 & 0.04 & 0.04 & 0.44 & 0.90 & 1.00\\
  & Oracle $(n)$ & 1.00 & 1.00 & 0.67 & 0.04 & 0.07 & 0.71 & 1.00 & 1.00\\ \hline
 Standard Error & Debiased SS & 0.32 & 0.32 & 0.31 & 0.31 & 0.31 & 0.31 & 0.32 & 0.32\\
  & MLE SS & 0.38 & 0.37 & 0.35 & 0.34 & 0.35 & 0.35 & 0.36 & 0.38\\
  & Debiased MS & 0.23 & 0.23 & 0.22 & 0.21 & 0.21 & 0.22 & 0.23 & 0.23\\
  & MLE MS & 0.27 & 0.26 & 0.25 & 0.25 & 0.25 & 0.26 & 0.26 & 0.26\\
  & Oracle $(n_2)$ & 0.34 & 0.31 & 0.30 & 0.31 & 0.31 & 0.31 & 0.32 & 0.33\\
  & Oracle $(n)$ & 0.23 & 0.22 & 0.21 & 0.21 & 0.21 & 0.21 & 0.22 & 0.23\\ \hline
 Empirical SD & Debiased SS & 0.36 & 0.37 & 0.34 & 0.33 & 0.35 & 0.35 & 0.39 & 0.38\\
  & MLE SS & 0.51 & 0.46 & 0.40 & 0.40 & 0.41 & 0.42 & 0.50 & 0.49\\
  & Debiased MS & 0.26 & 0.25 & 0.24 & 0.20 & 0.22 & 0.23 & 0.25 & 0.26\\
  & MLE MS & 0.35 & 0.32 & 0.29 & 0.24 & 0.26 & 0.27 & 0.31 & 0.34\\
  & Oracle $(n_2)$ & 0.32 & 0.33 & 0.33 & 0.29 & 0.30 & 0.34 & 0.32 & 0.34\\
  & Oracle $(n)$ & 0.24 & 0.23 & 0.24 & 0.20 & 0.23 & 0.22 & 0.24 & 0.24\\
  \hline
 \end{tabular*}}
 \bigskip
\end{table}

\end{document}